\documentclass[english,aps,prl,twocolumn,showpacs]{revtex4-1}
\usepackage[T1]{fontenc}
\usepackage[latin9]{inputenc}
\usepackage[active]{srcltx}
\usepackage{color}
\usepackage[colorlinks, citecolor=blue]{hyperref}
\usepackage{amsmath}
\usepackage{amssymb}
\usepackage{graphicx}

\makeatletter
\@ifundefined{textcolor}{}
{%
 \definecolor{BLACK}{gray}{0}
 \definecolor{WHITE}{gray}{1}
 \definecolor{RED}{rgb}{1,0,0}
 \definecolor{GREEN}{rgb}{0,1,0}
 \definecolor{BLUE}{rgb}{0,0,1}
 \definecolor{CYAN}{cmyk}{1,0,0,0}
 \definecolor{MAGENTA}{cmyk}{0,1,0,0}
 \definecolor{YELLOW}{cmyk}{0,0,1,0}
}

\makeatother

\usepackage{babel}
\begin{document}

\title{Electromagnetically induced transparency with controlled van der
Waals interaction}

\author{Huaizhi Wu, Meng-Meng Bian, Li-Tuo Shen, Rong-Xin Chen, Zhen-Biao
Yang, Shi-Biao Zheng}

\affiliation{Department of Physics, Fuzhou University, P. R. China}
\begin{abstract}
We study the electromagnetically
induced transparency (EIT) effect with two individually addressed
four-level Rydberg atoms subjected to the interatomic van der Waals
interaction. 
We derive an effectively atomic Raman transition model where two ladders of the usual Rydberg-EIT setting terminating at the same upper Rydberg level of long radiative lifetime are turned into a Rydberg-EIT lambda setup via two-photon transitions, leaving the middle levels of each ladder largely detuned from the coupling and probe laser beams. It can hence overcome the limits of applications for EIT with atoms of the ladder-type level configuration involving a strongly decaying intermediate state by inducing coherence between two ground states. 
By probing one of the atoms, we observe four doublets of absorption induced by the Autler-Townes (AT) splitting and the van der Waals interaction. 
In particular, we find
that the location of the EIT center remains unchanged compared
to the interatomic-interaction-free case, which demonstrated that the interference among the
multiple transition channels is basically destructive. The EIT with
controlled Rydberg-Rydberg interaction among few atoms provides a
versatile tool for engineering the propagation dynamics of light. 
\end{abstract}

\maketitle
Dipole-dipole or van der Waals (vdW) interaction between
atom pairs being excited to high-lying Rydberg states has great potential applications in quantum
information processing \cite{Saffman_RMP2010,Comparat_JOSAB2010},
many-body quantum simulation \cite{Weimer_2010NatPhy,Honer_PRL2010,Glaetzle_PRA2012,Honing_PRA2013,Olmos_PRL2013}
and nonlinear optics \cite{Peyronel_Nature2012,Parigi_PRL2012,Firstenberg_Nature2013}.
In the context of nonlinear optics, many efforts were dedicated to
map Rydberg-Rydberg interaction onto optical field using EIT \cite{Pritchard_PRL2010,Petrosyan_PRL2011,Gorshkov_PRL2011,Yan_PRA2012}.
The interatomic interaction within a Rydberg atomic vapour or cold
Rydberg gas can give rise to observable highly optical nonlinearity
\cite{Pritchard_PRL2010} and nonlocal optical effect \cite{Sevincli_PRL2011}
as well as new type of photonic quantum gas \cite{Otterbach_PRL2013}
under conditions of EIT. Moreover, the enhanced optical nonlinearity
promises engineering of photonic dissipative many-body dynamics \cite{Gorshkov_PRL2013}
and observation of photon-photon interaction towards the few photons
regime \cite{Gorshkov_PRL2011,Shahmoon_PRA2011,He2014}. The
interatomic interaction modified photon correlation can also exhibit a significant
back action on the Rydberg atom statistics \cite{Hofmann_PRL2013}.
Most of the EIT schemes involved Rydberg atoms with the ladder-typed
level structure, where the coherence is induced between a ground state
and a Rydberg state via a  strongly decaying intermediate state \cite{Peyronel_Nature2012,Parigi_PRL2012,Firstenberg_Nature2013,Pritchard_PRL2010,Petrosyan_PRL2011,Gorshkov_PRL2011,Gorshkov_PRL2013,Shahmoon_PRA2011,Hofmann_PRL2013,Yan_PRA2012,Otterbach_PRL2013,Sevincli_PRL2011,He2014}. 

Besides the studies towards controlling the propagation dynamics,
the observation and mastering of the coherent dynamics among few Rydberg
atoms in the limit of strong (or partial) blockade has been
an essential goal for the dipole blockade based quantum information
processing \cite{Saffman_RMP2010,Comparat_JOSAB2010,knoernschild_APL2010}.
Remarkable experimental advance has been recently made for the realization
of two-qubit quantum entanglement \cite{Wilk_PRL2010,Beguin_PRL2013}
and quantum gates \cite{Isenhower_PRL2010,Zhang_PRA10_esCNOT} via
join or individual laser steering. In particular, by controlling the
interatomic distance $R$, the characteristic $C_{6}/R^{6}$ dependence
of the vdW interaction strength between two interacting Rydberg atoms has been observed directly by tracking the time-dependent collectively coherent dynamics \cite{Beguin_PRL2013}.
It thus becomes very promising for studying quantum effects and nonlinear
optical phenomena with well-localized Rydberg atoms prepared in well
defined quantum state \cite{Weidemuller_Physics2013,Browaeys2014,Goerz2014}.

\begin{figure}
\includegraphics[width=1\columnwidth]{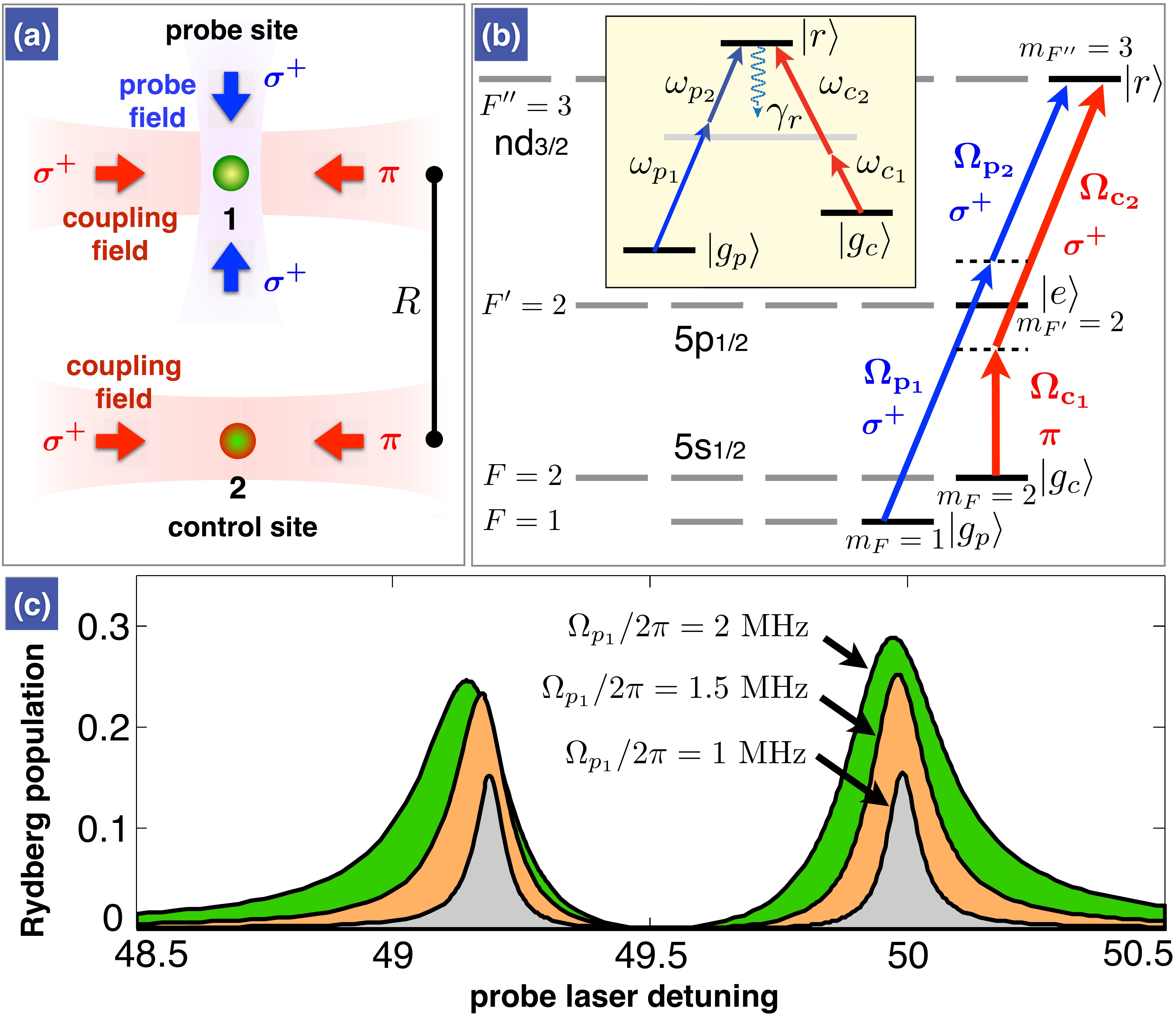}

\caption{\label{fig:setup&scheme}(Color
 online) (a) Sketch of setup. (b) Atomic level structure
and laser addressing scheme. The level structure are the Zeeman substates
of the two hyperfine ground states $F=1,2$ of $5s_{1/2}$, the excited
state $F'=2$ of $5p_{1/2}$ and $F''=3$ of the high-lying Rydberg
state $nd_{3/2}$. A $\pi$ polarized laser beam red-detuned to the
$|F=2,m_{F}=2\rangle\rightarrow|F'=2,m_{F'}=2\rangle$ transition
and a $\sigma^{+}$ polarized laser beam blue-detuned to the $|F'=2,m_{F'}=2\rangle\rightarrow|F''=3,m_{F''}=3\rangle$
transition together act as the coupling channel. While the probe scan
is implemented by two $\sigma^{+}$ polarized laser beams blue-detuned
to the $|F=1,m_{F}=1\rangle\rightarrow|F'=2,m_{F'}=2\rangle$ transition
and red-detuned to the $|F'=2,m_{F'}=2\rangle\rightarrow|F''=3,m_{F''}=3\rangle$
transition, respectively. The inset shows the reduced effective model where the two ladders of the usual Rydberg-EIT setting terminating at the same upper Rydberg level are turned into the usual EIT lambda setup after eliminating the intermediate level  $|e\rangle$ for the large detuning regime. (c) The Rydberg spectra without interatomic
interaction for increasing probe Rabi frequencies $\Omega_{p_{1},p_{2}}$.
The resonance peaks are displaced due to the Stark shifts induced
by the coupling and probe laser beams. Parameters are $\Delta_{p_{2}}/2\pi=50$
MHz, $\Omega_{p_{2}}=\Omega_{p_{1}}$, $\Delta_{c_{1},c_{2}}/2\pi=1$ GHz, $\Omega_{c_{1},c_{2}}/2\pi= 20$ MHz, and $(\gamma_{ec},\gamma_{ep},\gamma_{r})/2\pi=(3,3,0.1)$ MHz. }
\end{figure}

Here we propose a nonlocal EIT scheme with two four-level Rydberg
atoms subjected to the vdW interaction. In our model, two ladders of the usual Rydberg-EIT setting terminating at the same upper Rydberg level are turned into the usual EIT lambda setup with the highest level (of long radiative lifetime) being the most strongly decaying one (see Fig.\ref{fig:setup&scheme}). This was realized by using two-photon transitions where the laser beams are tuned largely detuned from the intermediate level. It hence allows for observing a dissipation-insensitive coherence induced between two ground states \cite{Harris_PRL1990_EIT},
and overcomes the limits of the ladder configuration for
applications where the strongly decaying intermediate state is highly involved and a (meta)stable dark state is absent for two interacting atoms  in the strict sense \cite{Fleischhauer_RMP2005,Schempp_PRL2010}. By probing the ground-Rydberg transition channel under EIT for one atom, we find that the vdW interaction strength
can be directly mapped onto the optical absorption spectra
consisting of four absorption doublets jointly induced by the interatomic
interaction and the AT splitting. The absorption spectra display a very striking feature
that the location of the EIT center remains unchanged compared with
the single-atom EIT scheme as the interatomic interaction energy increases,
which stems from the destructive quantum interference among all the
transition channels.

The setup for the Rydberg-EIT proposal, as schematically shown in
Fig.\ref{fig:setup&scheme}(a), involves two $^{87}$Rb atoms located
in separated optical traps (denoted as control site and probe site)
giving access to individual laser addressing \cite{Zhang_PRA10_esCNOT,Beguin_PRL2013}.
The coupling field, consisting of two laser beams with $\pi$
and $\sigma^{+}$ polarization respectively, irradiates the pair of
atoms and strongly drives the atomic transition from the ground state
$|g_{c}\rangle$ to the high-lying Rydberg
level $|r\rangle$ mediated by the
excited state $|e\rangle$. The EIT
spectroscopy is probed with another two $\sigma^{+}$ polarized laser
beams just shining on the probe site and scanning across the $|g_{p}\rangle\rightarrow|r\rangle$
atomic resonance while the $C_{6}/R^{6}$ dependent interatomic vdW interaction emerges (see figure \ref{fig:setup&scheme} caption  
for details) \cite{Beguin_PRL2013}.\textcolor{black}{{} }The spontaneous
decay rates for the atomic transitions $|r\rangle\rightarrow|e\rangle$,
$|e\rangle\rightarrow|g_{c}\rangle$ and $|e\rangle\rightarrow|g_{p}\rangle$,
are $\gamma_{r}$, $\gamma_{ec}$ and $\gamma_{ep}$, respectively.
\textcolor{black}{Note that the probability amplitude for finding
the atom in the Rydberg state $|r\rangle$ can be used to evaluate
the imaginary part of the linear susceptibility for optical response
\cite{Ates_PRA2011}.}\textcolor{green}{{} }Therefore, the Rydberg spectra,
i.e. the Rydberg population (corresponding to the probability for two-photon absorption) for the atom at the probe site versus the probe laser detuning from two-photon resonance will be detailedly studied here \cite{Agarwal_PRL1996}.

In the rotating frame of the atomic bare energies ($\omega_{g_{p}}$, $\omega_{g_{c}}$, $\omega_{e}$ and $\omega_{r}$), the dynamics of
the two-atom system is governed by the Hamiltonian (set $\hbar=1$)
\begin{eqnarray}
\mathcal{H} & =& \sum_{l=1,2}(\Omega_{c_{1}}e^{-i\Delta_{c_{1}}t}|e\rangle_{ll}\langle g_{c}|+\Omega_{c2}^{*}e^{-i\Delta_{c_{2}}t}|e\rangle_{ll}\langle r|+H.c.)\nonumber \\
 & + &
 (\Omega_{p_{1}}^{*}e^{-i\Delta_{p_{1}}t}|g_{p}\rangle_{11}\langle e|+\Omega_{p_{2}}e^{-i\Delta_{p_{2}}t}|r\rangle_{11}\langle e|+H.c.)\nonumber \\
 & + & \mathcal{V}(R)|r\rangle_{1}|r\rangle_{22}\langle r|_{1}\langle r|
\end{eqnarray}
with the attractive potential
$\mathcal{V}(R)=-C_{6}/R^{6}$ ($C_{6}>0$)  \cite{Singer_JPB2004,Reinhard_PRA2007}. Here $\Delta_{p_{1}}=\omega_{e}-\omega_{g_{p}}-\omega_{p_{1}},$
$\Delta_{p_{2}}=\omega_{p_{2}}-\omega_{r}+\omega_{e},$ $\Delta_{c_{1}}=\omega_{c_{1}}-\omega_{e}-\omega_{g_{c}},$
$\Delta_{c_{2}}=\omega_{r}-\omega_{e}-\omega_{c_{2}}$, $\omega_{y}$ and $\Omega_{y}$
($y=p_{1},p_{2},c_{1},c_{2}$) are laser frequencies and atom-laser coupling strengths for the related atomic transitions as indicated in the Fig.\ref{fig:setup&scheme}(b).
The atoms are assumed to identically interact with the coupling laser
field. For the dispersive regime $\Omega_{y}\ll\Delta_{y}$, defining the \textit{\textcolor{black}{two-photon probe laser detuning}}
${\color{black}\delta_{p}=(\omega_{p_{1}}+\omega_{p_{2}})-(\omega_{r}-\omega_{g_{p}})}$ and coupling laser detuning
$\delta_{c}=(\omega_{c_{1}}+\omega_{c_{2}})-(\omega_{r}-\omega_{g_{c}})$,
and using the time-averaging method \cite{James_effectiveH}, we finally
pass to a new interaction Hamiltonian in the rotating frame with respect
to $\mathcal{H}_{0}=\sum_{l=1,2}\sum_{x=g_{p},g_{c},r}\mu_{l,x}|x\rangle_{ll}\langle x|$
(with $\mu_{1,g_{p}}=\delta_{p}+\alpha_{1}$, $\mu{}_{2,g_{p}}=0$, $\mu_{l,g_{c}}=-\delta_{c}+\alpha_{l}$,
$\mu_{l,r}=\alpha_{l}$, $\alpha_{1}=\frac{|\Omega_{c_{2}}|^{2}}{\Delta_{c_{2}}}-\frac{|\Omega_{p_{2}}|^{2}}{\Delta_{p_{2}}}$,
and $\alpha_{2}=\frac{|\Omega_{c_{2}}|^{2}}{\Delta_{c_{2}}}$) 
\begin{eqnarray}
\mathcal{H}_{e} & = & \sum_{k={c,p}} \varepsilon_{k}|g_{k}\rangle_{11}\langle g_{k}|+\varepsilon'_{c}|g_{c}\rangle_{22}\langle g_{c}|+( \sum_{k={c,p}}  \lambda_{k}|r\rangle_{11}\langle g_{k}| \nonumber \\
 & + &\lambda_{c}|r\rangle_{22}\langle g_{c}|+H.c.)+ \mathcal{V}(R)|r\rangle_{1}|r\rangle_{22}\langle r|_{1}\langle r|, \label{eq:He}
\end{eqnarray}
where $\lambda_{k}=-\frac{1}{2}\Omega_{k_{1}}^{*}\Omega_{k_{2}}(\frac{1}{\Delta_{k_{1}}}+\frac{1}{\Delta_{k_{2}}})$,
$\varepsilon_{k}=\beta_{k}-\mu_{1,g_{k}}$, 
$\varepsilon'_{c}=\beta_{c}-\mu_{2,g_{c}}$, 
$\beta_{p}=-\frac{|\Omega_{p_{1}}|^{2}}{\Delta_{p_{1}}}$ and
$\beta_{c}=\frac{|\Omega_{c_{1}}|^{2}}{\Delta_{c_{1}}}$.
The system is now described by the time independent interaction Hamiltonian
corresponding to an effective three-level Raman (TLR) model, which
has been verified numerically to be effective. Since the atoms interact dispersively with the laser beams,
the population of the intermediate state $|e\rangle$ approximates
$\sum_{y}(\Omega_{y}/\Delta_{y})^{2}$ and can be ignored without
considering dissipation.

The density operator $\rho$ for the two-atom system {[}Eq. (\ref{eq:He}){]}
can be expressed by a $9\times9$ matrix. For a given set of parameters,
the full dynamics including atomic spontaneous emission is calculated
from the master equation with the Lindblad form
$
\dot{\rho}  = -i[\mathcal{H}_e,\rho]+\sum_{j=1,2}\mathcal{L}_{j},
$
where
$
\mathcal{L}_{j}=\frac{1}{2}\sum_{k=c,p}\Gamma_{rk}^{(j)}(2\sigma_{kr}^{(j)}\rho\sigma_{rk}^{(j)}-\sigma_{rk}^{(j)}\sigma_{kr}^{(j)}\rho-\rho\sigma_{rk}^{(j)}\sigma_{kr}^{(j)})
$,
$\sigma_{kr}^{(j)}=|g_k\rangle_{jj}\langle r|$, and
$\Gamma_{rk}^{(j)}$
are effective spontaneous decay rates from the Rydberg level $|r\rangle$
to the ground states and are given by 
$\Gamma_{rk}^{(1)}\approx\gamma_{ek}\gamma_{r}/(\gamma_{ec}+\gamma_{ep}+\gamma_{r})$,
$\Gamma_{rp}^{(2)}=0$ and $\Gamma_{rc}^{(2)}\approx(\gamma_{ec}+\gamma_{ep})\gamma_{r}/(\gamma_{ec}+\gamma_{ep}+\gamma_{r})$.
Here we have considered the fact that the population
of the intermediate excited state $|e\rangle$ (far off-resonantly
coupling to the laser beams) induced by the spontaneous emission of
the Rydberg state $|r\rangle$ will quickly decay to the ground states
for $(\gamma_{ec},\gamma_{ep})\gg\gamma_{r}$ \cite{Beguin_PRL2013}.
Thus, the time scale for the dissipative transitions from the Rydberg
state to the ground states is dominated by $1/\gamma_{r}$. We neglect
dephasing due to atomic collisions as well as amplitude and phase
fluctuations of the laser beams, and focus on the effect induced
by the vdW interaction. While the probe laser beams are applied, steady
states can be found by solving the master equation numerically. Then,
the Rydberg spectra is calculated by summing over all the diagonal
elements involving the Rydberg population of the probed atom.

\begin{figure}
\includegraphics[width=1\columnwidth]{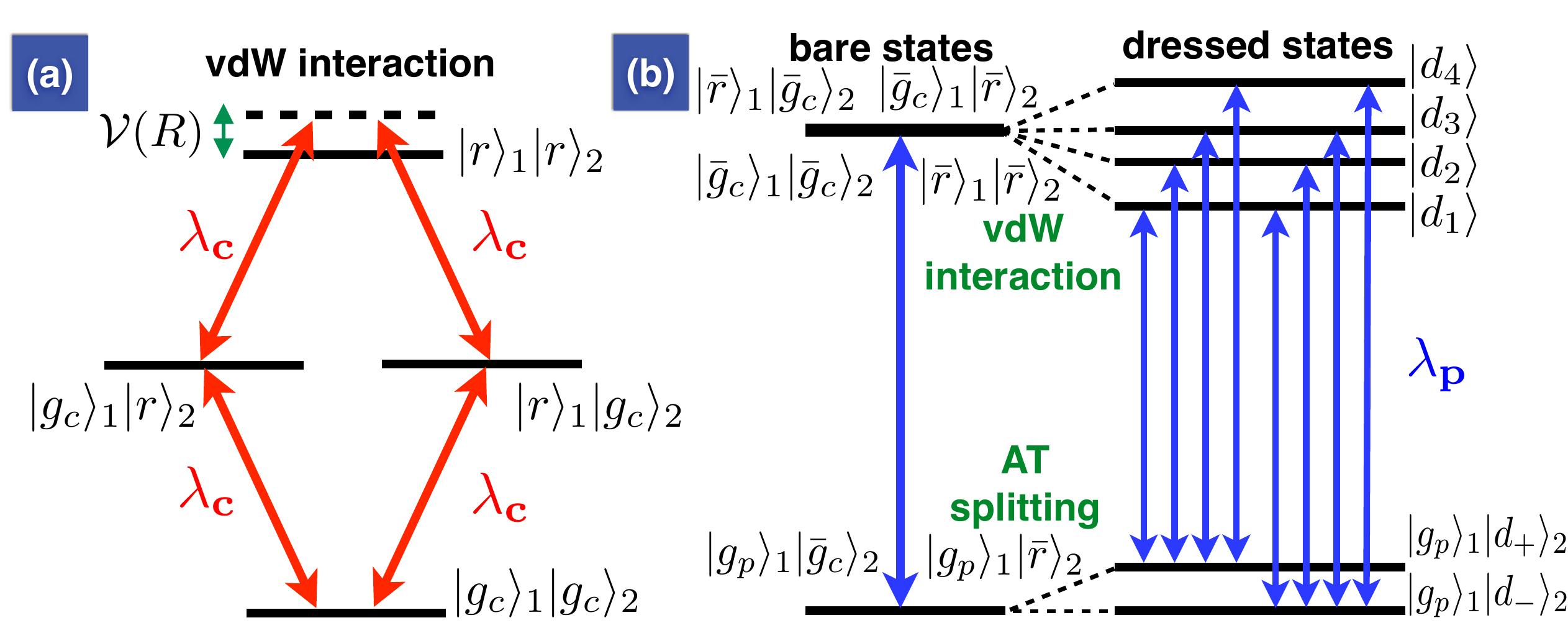}

\caption{\label{fig:vdW_esplitting}(Color
 online) (a) Two-atom transition scheme involving
the vdW interaction. The collective Rydberg excited state is shifted
according to the attractive potential $\mathcal{V}(R)$. (b) Atomic
transition channels induced by the vdW interaction and the AT splitting, not to scale (see text).}
\end{figure}

(a) The Rydberg spectra without interparticle interaction. We first
displace the coupling laser beams away from the control atom \cite{Urban_NatPhy09},
then the atoms are free to the vdW interaction and are
uncorrelated. It corresponds to a single atom EIT scheme involving
a $\Lambda$-type three-level system, which is simply described by
the Hamiltonian 
\begin{eqnarray}
\mathcal{H}_{e}^{nv} =  \sum_{k={c,p}} \varepsilon_{k}|g_{k}\rangle_{11}\langle g_{k}| +(\lambda_{k}|r\rangle_{11}\langle g_{k}|+H.c.).\label{eq:H_single_atom}
\end{eqnarray}
The strong non-perturbative coupling creates two dressed states (eigenstates)
for $\mathcal{H}_{e}^{nv}$ given by 
$
|d_{\pm}^{nv}\rangle_{1}=[(\varepsilon_{c}/2\pm\varpi)|\bar{g}_{c}\rangle_{1}+\lambda_{c}|\bar{r}\rangle_{1}]/\mathcal{N}_{\pm},\label{eq:1at_dressed}
$
with $\varpi=\sqrt{\varepsilon_{c}^{2}/4+\lambda_{c}^{2}}$, $\mathcal{N}_{\pm}$ the normalization factors, and the relative
eigenenergies $E_{\pm}^{nv}=\varepsilon_{c}/2\pm \varpi$.
Here, we have denoted the system states by $|\bar{\varphi}\rangle$
($\varphi=g_{c},r$), which are product states of the 'bare' atom
and the coupling laser field for clarity. Note that the
states of the 'bare' atom are now dressed by the resonant laser field
through the AC Stark effect and are separated by $2\varpi$ (known as AT splitting) \cite{AT_PR1955}.

In the limit of perturbative probe (i.e. $\lambda_{p}\ll\lambda_{c}$),
the atomic transitions $|g_{p}\rangle_{1}\rightarrow|d_{\pm}^{nv}\rangle_{1}$
are in resonance while the condition $\varepsilon_{p}=E_{\pm}^{nv}$
is fulfilled. Thus, the location of the AT doublet corresponding
to two-photon resonant absorption lines are found at the probe detunings
\begin{equation}
\delta_{p}^{(\pm)}(\omega_{p}=\omega_{p_1}+\omega_{p_2})=E_{\pm}^{nv}+E_{ss},\label{eq:EIT_resonance}
\end{equation}
contributed together by the Stark shifts $E_{ss}=(\beta_{p}-\alpha_1)$ induced by the probe
and coupling field. This is verified by the Rydberg spectra shown in the Fig.\ref{fig:setup&scheme}(c). 

The EIT resonance condition {[}the Eq.(\ref{eq:EIT_resonance}){]}
can be derived alternatively from the imaginary part of the linear
susceptibility calculated with the Hamiltonian $\mathcal{H}_{e}^{nv}$
and keeping track of the terms that oscillate with the frequency $\omega_{p}$
\cite{Fleischhauer_RMP2005},
\begin{equation}
\textrm{Im}[\chi^{(1)}]\propto\frac{(\varepsilon_{p}-\varepsilon_{c})\varepsilon_{c}\gamma_{r}}{|\varepsilon_{p}(\varepsilon_{p}-\varepsilon_{c})-\lambda_{c}^{2}-i\frac{1}{2}\gamma_{r}\varepsilon_{c}|^{2}}.
\end{equation}
Indeed, the effect of the total decoherence rates (including the
atomic spontaneous emission and the finite linewidth of the laser
induced dephasing) on $\textrm{Im}[\chi^{(1)}]$ can be considered
simply by replacing $\gamma_{r}$ by total decaying rate \cite{Fleischhauer_RMP2005}.
Compared with the original EIT scheme \cite{Harris_PRL1990_EIT},
the profile of the spectra here are essentially displaced due to the
Stark shift $\alpha_2$  of the Rydberg state mainly induced by the coupling laser
field for weak probe. No broadening
of the lines was observed for increasing probe Rabi frequencies {[}see
Fig.\ref{fig:setup&scheme}(c){]}.

(b) The Rydberg spectra with controlled vdW interaction. While the
coupling laser beams irradiate both control and probe sites, the potential
admixture of the Rydberg state $|r\rangle$ with $|g_{c}\rangle$
for both atoms will include the collective Rydberg excitation giving
rise to the long range vdW interaction and the partial blockade of
the Rydberg biexcitation, see figure \ref{fig:vdW_esplitting}(a). Inspired
by the case without interparticle interaction, we will adopt the dressed
atom picture to gain insight into the dipolar interaction involved
EIT phenomenon. 

Again for $\lambda_{p}\ll\lambda_{c}$, we first concentrate on the
coupling channels, in which case the system Hamiltonian narrated by
the TLR model reduces to 
\begin{eqnarray}
\mathcal{H}_{c}^{v} & = & \varepsilon_{c}|g_{c}\rangle_{11}\langle g_{c}|+\varepsilon'_{c}|g_{c}\rangle_{22}\langle g_{c}|+\underset{l=1,2}{\sum(}\lambda_{c}|r\rangle_{ll}\langle g_{c}|+H.c.)\nonumber \\
 & + & \mathcal{V}(R)|r\rangle_{1}|r\rangle_{22}\langle r|_{1}\langle r|.
\end{eqnarray}
The temporal evolution of the system remains in the subspace (written
in new notation) spanned by 
$
\{|\bar{\phi} _{n=1,2,3,4} \rangle\}
$
with $|\bar{\phi}_1\rangle=|\bar{g}_{c}\rangle_{1}|\bar{g}_{c}\rangle_{2}$, $|\bar{\phi}_2\rangle=|\bar{g}_{c}\rangle_{1}|\bar{r}\rangle_{2}$, $|\bar{\phi}_3\rangle=|\bar{r}\rangle_{1}|\bar{g}_{c}\rangle_{2}$, $|\bar{\phi}_4\rangle=|\bar{r}\rangle_{1}|\bar{r}\rangle_{2}$.
It is readily to obtain the dressed states (i.e. the eigenstates for
$H_{c}^{v}$) for the interacting two-atom system that can be expressed
by
$
|d_{j}^{v}\rangle  = \sum_n \eta_{n,j}|\bar{\phi}_n\rangle
$
$(j=1,2,3,4)$, and the corresponding dressed energies $E_{j}^{v}$.
Note that the new dressed levels $\{|d_{j}^{v}\rangle\}$ have been
shifted by $\langle d_{j}^{v}|\mathcal{V}(R)|\bar{r}\rangle_{1}|\bar{r}\rangle_{22}\langle\bar{r}|_{1}\langle\bar{r}|d_{j}^{v}\rangle=|\eta_{4,j}|^{2}\mathcal{V}(R)$
due to the vdW interaction,
and will serve as the collective excited states. On the other hand, the probe atom is initially in $|g_{p}\rangle_{1}$,
and the control atom is continuously shined by the coupling laser
beams, which creates two collective ground states $|g_{c}\rangle_{1}|d_{\pm}^{nv}\rangle_{2}$
with the separation given by the AT splitting for the initially non-interacting
system in the steady state. Hence, the probe laser will scan across
the transition channels among the two splitted collective ground states
and the four collective excited states. We then observe eight
resonant absorption peaks arising at the probe detunings divided by
two groups
\begin{equation}
\delta_{p}^{(\pm j)}(\omega_{p})=E_{j}^{v}-E_{\pm}^{nv}+E_{ss},
\end{equation}
where $|\delta_{p}^{(+j)}-\delta_{p}^{(-j)}|=|E_{+}^{nv}-E_{-}^{nv}|$
are the AT splittings. For simplicity, we can further assume $\Omega_{c_{1}}=\Omega_{c_{2}}$
and $\Delta_{c_{1}}=\Delta_{c_{2}}$, and then have $\varepsilon_{c}=\varepsilon'_{c}=0$
without probe scan. In this case, the system states $\{|\bar{\phi} _{n} \rangle\}$
are degenerate without the interparticle interaction and the eigenstates
for the Hamiltonian involving only the coupling laser field are just
the product of the single-atom dressed states $|d_{\pm}^{nv}\rangle_{l}$
($l=1,2$). Only when the vdW interaction
is included will the atoms become correlated. Then, the degeneracy is broken and the dressed states $\{|d_{j}^{v}\rangle\}$
including an unshifted dark state $(|\bar{g}_{c}\rangle_{1}|\bar{r}\rangle_{2}-|\bar{r}\rangle_{1}|\bar{g}_{c}\rangle)/\sqrt{2}$
are found. The two-atom probe scheme in this limit is sketched in Fig. \ref{fig:vdW_esplitting}(b).

\begin{figure}
\includegraphics[width=1\columnwidth]{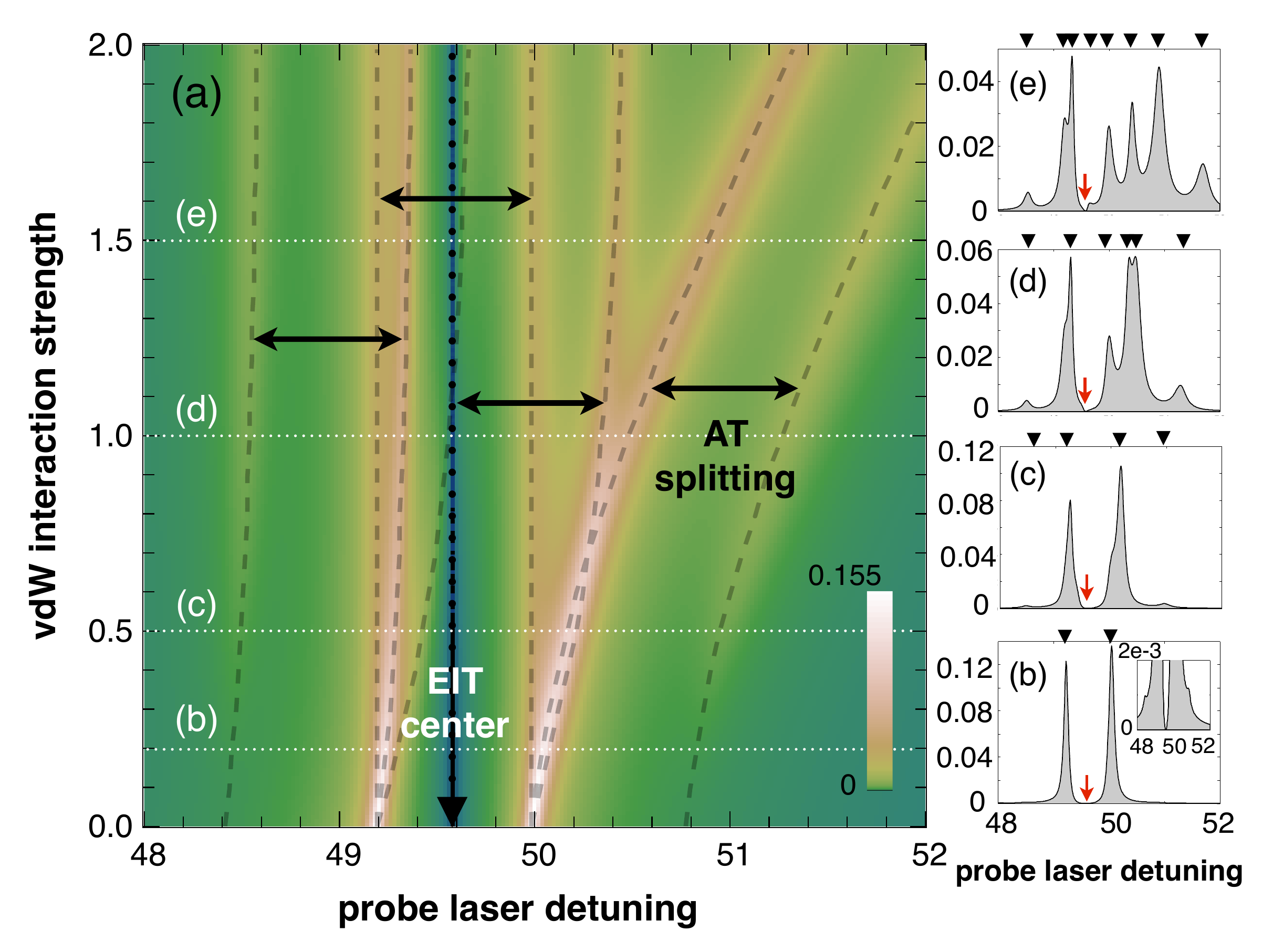}
\caption{\label{fig:spectra_vdW}(Color
 online) (a) Rydberg population versus probe laser
detuning and the vdW interaction with $\Omega_{p_{1},p_{2}}/2\pi=1$ MHz, $\Delta_{p_{2}}/2\pi=50$ MHz; further parameters are the same as in Fig.\ref{fig:setup&scheme}(c) and we have assumed the two atoms uniformly interact with the coupling laser beams. As predicted in Fig.\ref{fig:vdW_esplitting}(b), eight
resonant peaks (corresponding to the eight dipole-allowed atomic transition
channels) can be resolved in the Rydberg spectra as the vdW interaction
strength increases. A detailed calculation by considering two interacting
Rydberg atoms with one of them being probed fits well with the numerical
results (black dash line). (b)-(e) Linecut of the Rydberg
spectra along $\mathcal{V}(R)/2\pi=0.2$, 0.5, 1, and 1.5
MHz in (a). The EIT center and the absorption peaks are
indicated.}
\end{figure}

The Rydberg spectra involving the interparticle
vdW interaction is presented in Fig. \ref{fig:spectra_vdW}. As the interaction strength increases, all the transition channels
get more involved and we can resolve the resonance peaks more
clearly; the peak value of the Rydberg spectra reduces accordingly
as the blockade effect enhances. All the eight peaks predicted before
can be seen at the probe detunings given by $\delta_{p}^{(\pm j)}(\omega_{p})$
while $\mathcal{V}(R)$ becomes comparable with $|E_{+}^{nv}-E_{-}^{nv}|$.
In between the peaks, there exists nonvanishing Rydberg population
except for $\delta_{p}=E_{ss}+(E_{+}^{nv}+E_{-}^{nv})/2$,
which is exactly the location of the EIT center found in the interatomic
interaction free scheme {[}see Fig.  \ref{fig:spectra_vdW}(b)-(e){]}. It is striking that
the transparency can be observed\textit{ in situ} even though the
vdW interaction is significant for the two-body dynamics, which then
implies that the quantum interference among the multiple transition
channels must be destructive. The peaks distribute asymmetrically
around the EIT center. Moreover, the extended profile of
the Rydberg spectra allows one to control the absorption of light
in a wider range of laser frequency compared with the case without
interatomic interaction. 

\begin{figure}
\includegraphics[width=1\columnwidth]{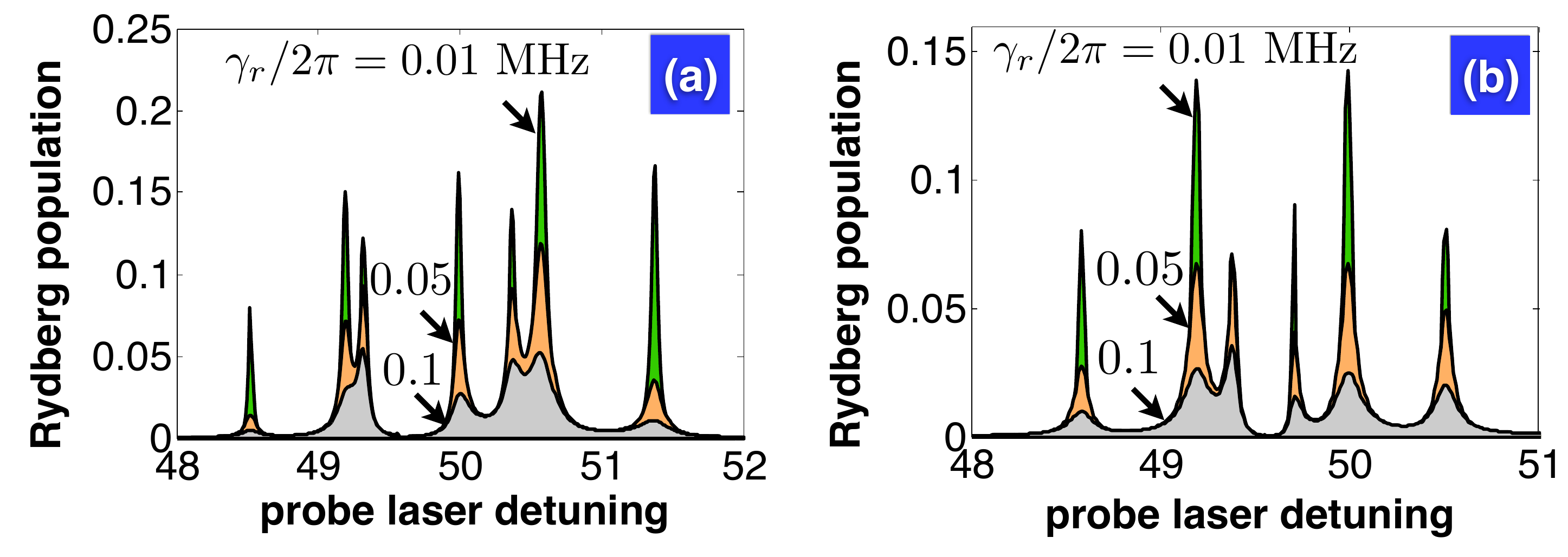}

\caption{\label{fig:spectra_gamma}(Color
 online) Rydberg spectra for distinct spontaneous emission rates $\gamma_{r}/2\pi=0.01$,
$0.05$, $0.1$ MHz, corresponding to different Rydberg levels. The
interatomic interaction strengths are (a) $\mathcal{V}(R)/2\pi=1.1$
MHz and (b) $\mathcal{V}(R)/2\pi=4$ MHz, respectively. Further parameters as in Fig.\ref{fig:spectra_vdW}.}
\end{figure}

To carry out the proposed EIT scheme experimentally, the atoms must be close
enough to experience the vdW interaction, yet far enough apart that
one of the atoms can be probed. This may be
implemented with the new developed setup used for direct measurement
of the vdW interaction between two Rydberg atoms \cite{Beguin_PRL2013},
where the interaction strength ranged from 0 to 10 MHz can be flexibly
tuned by controlling the interatomic separation of a few micrometers.
It thus features individual addressing of the atoms by applying
the optical-wavelength laser beams focused to a small waist \cite{Urban_NatPhy09,Browaeys_PRA2010_CohExc}.
On the other hand, the Rydberg level $|r\rangle$ with the principal
quantum number $n$ has the radiative lifetime scaling as $n^{3}$.
Therefore, the spontaneous emission rate $\gamma_{r}$ of $|r\rangle$
should be much less than that of the intermediate level $|e\rangle$
($\sim2\pi\times3$ MHz) for large $n$. We have shown the Rydberg
spectra for varied $\gamma_{r}$ (corresponding to different Rydberg
states) in figure \ref{fig:spectra_gamma}, which depicts that the unshifted resonance peaks
can be better resolved as $\gamma_{r}$ decreases.

In conclusion, we have studied the Rydberg-EIT phenomena with two interacting Rydberg atoms effectively described by the TLR model where the atomic middle levels are off-resonantly eliminated via two-photon transitions. An unshifted EIT center is found due to destructive interference of the multiple transition channels originated from the vdW interaction and the AT splitting. The nonlocal Rydberg-EIT may enable flexible modulation of light propagation and better understanding of the linear or nonlinear optical response
with few photons and few atoms, e.g., by resorting to microresonator enhanced precision measurement used for realizing photon turnstile  \cite{Dayan_Science2008} and optical switch \cite{Shea_PRL2013} with a single atom.

This work was supported by the '973'
Program of China under Grant No. 2012CB921601, the NSF of China under Grant Nos. 11305037, 11347114 and 11374054, the NSF of Fujian Province
under Grant No. 2013J01012, and the fund from Fuzhou University.

\bibliographystyle{apsrev}
\bibliography{/Users/huaizhiwu/Lab_Huaizhi/Bibtex/EIT,/Users/huaizhiwu/Lab_Huaizhi/Research_Projects/EIT_Y_atoms/reference/EIT_Rydberg}

\end{document}